\documentclass[manuscript,screen,nonacm]{acmart}

\usepackage{tabularx}
\usepackage{soul}

\setcopyright{none}
\renewcommand\footnotetextcopyrightpermission[1]{}

\AtBeginDocument{%
  }

\begin{document}

\title{How Generative AI Redefines Personalization and Diversity in the Newsroom}

\author{Cheng Yu}
\orcid{0009-0001-0102-9441}
\affiliation{%
  \institution{Technical University of Munich}
  \country{Germany}
}

\author{Jim Chi}
\orcid{0009-0001-6852-0999}
\affiliation{%
  \institution{Queensland University of Technology}
  \country{Australia}
}

\author{Aaron J. Snoswell}
\orcid{0000-0003-0960-0144}
\affiliation{%
  \institution{Queensland University of Technology}
  \country{Australia}
}

\author{Orestis Papakyriakopoulos}
\orcid{0000-0003-4680-0022}
\affiliation{%
  \institution{Technical University of Munich}
  \country{Germany}
}

\renewcommand{\shortauthors}{Yu et al.}

\title{How Generative AI Redefines Personalization and Diversity in the Newsroom}

\begin{abstract}
The tension between news personalization and information diversity has traditionally been understood as a matter of different content reaching different people, with diversity evaluated by the range of sources, topics, and perspectives to which audiences are exposed. 
Generative AI outdates this understanding, since it transforms personalization in two ways: how journalists adapt news to audiences and how the AI tools journalists use personalize content production to their own preferences. 
Drawing on 16 semi-structured interviews with media professionals across public service, commercial, and state-affiliated news settings, we examine how these practices reshape editorial judgment and the conditions under which news diversity is produced. 
We find that participants negotiate AI integration by maintaining authority over meaning while delegating aspects of content presentation. Yet the tools they adapt for their own work also mediate which sources and perspectives become available, introducing diversity-relevant decisions that remain largely invisible to organizational oversight. 
While participants anticipate diversity risks in audience-facing personalization, they predominantly interpret diversity-imposed limitations in their own tools as technical problems. 
As a result, decisions that shape news diversity increasingly occur before publication and across multiple versions of a story, where neither evaluation of published items nor human approval can readily observe them. Journalists thus remain formally responsible for outcomes shaped by decisions they cannot fully see or control.
Our findings highlight that safeguarding news diversity requires looking beyond published content to the professional practices, infrastructures, and distributions of control that shape it.
\end{abstract}

\ccsdesc[500]{Human-centered computing~Empirical studies in collaborative and social computing}

\keywords{generative AI, news personalization, diversity, editorial judgment, professional jurisdiction}


\maketitle

\section{Introduction}
Personalization and diversity are central to how news media serve the public, yet balancing them is a longstanding challenge~\cite{vandenbulck2018}. 
Personalization can make news more relevant and accessible to individual audiences, while tailoring news to individual preferences can narrow exposure to information beyond those preferences and fragment shared public knowledge~\cite{helberger2018}. 
Research on digital news personalization has primarily examined how news services select and distribute existing stories~\cite{thurman2012,bodo2019}. 
Drawing on Napoli's distinction between source diversity (who produces news), content diversity (what is produced), and exposure diversity (what audiences encounter)~\cite{napoli1999}, much of this work addresses exposure diversity: whether recommendations broaden or narrow what audiences encounter~\cite{moller2018}, how diversity can be measured in recommendations~\cite{vrijenhoek2021}, whether personalization fragments publics~\cite{nechushtai2019}. 
These approaches generally take the news item as its unit of analysis, assuming a relatively stable piece of journalistic content that can be selected and distributed to different audiences.

Generative AI (GenAI) challenges the assumption of a relatively stable news item as the unit of analysis. News personalization no longer involves only selecting which stories reach different audiences; the stories themselves can also be adapted. A single report can take different lengths, languages, formats, and emphases for different audiences, potentially changing which facts, contexts, and perspectives each reader encounters. 
At the same time, GenAI expands where personalization can occur. Journalists and other newsroom practitioners can configure general-purpose AI tools to suit their professional needs and workflows. As these individually adapted tools support research, sourcing, and drafting, they may shape which sources, information, and perspectives practitioners encounter and draw upon in producing news. 
We conceptualize these developments as two-way personalization: 
\textit{tool personalization}, which shapes the information and tools through which reporting is produced and \textit{audience personalization}, which shapes how reporting reaches different readers.

GenAI does not merely introduce new biases into journalism; it changes where diversity-relevant decisions occur and whether conventional evaluation and oversight can observe them. 
Existing approaches to news diversity have provided important ways of examining who produces news~\cite{voakes1996}, what information is published~\cite{masini2018,loecherbach2020}, and which news items audiences encounter~\cite{helberger2018,moller2018,vrijenhoek2021}.
However, evaluating published output alone overlooks two consequential forms of AI mediation. Before publication, individually configured AI tools can shape which sources and perspectives become available for editorial consideration, leaving relevant omissions invisible in the resulting report. After publication, generative adaptation can produce multiple versions of the same story that retain its overall subject while differing in essential claims, evidential qualifications, or context. These extend familiar concerns about source bias and inaccurate rewriting, but they arise where conventional evaluation does not look. They reveal the limits of treating finished news items as sufficient objects of diversity evaluation, and human approval as sufficient evidence of editorial control. Governance increasingly rests on the latter: the EU AI Act, for example, exempts AI-generated public-interest text from disclosure when it has undergone human review or editorial control and a person holds editorial responsibility~\cite{aiact2024}. Therefore evaluating and governing news diversity under GenAI requires attention to the informational conditions of news production and to the transformations through which reported information reaches different audiences.
We ask: 

\textbf{RQ1.} How is generative AI incorporated into journalistic work, and how do existing professional norms, working practices, and organizational arrangements afford its integration?

\textbf{RQ2.} How is news diversity considered and evaluated across audience-facing personalization and practitioners' use of generative AI?

We answer these questions with 16 semi-structured interviews with editorial, technical, product, strategy, and
legal professionals in public service, commercial, and state-affiliated news settings across Europe and the Asia-Pacific,
complemented by two targeted validation studies of model behavior. We make the following contributions:

\begin{figure}[t]
  \centering
  \includegraphics[width=\linewidth]{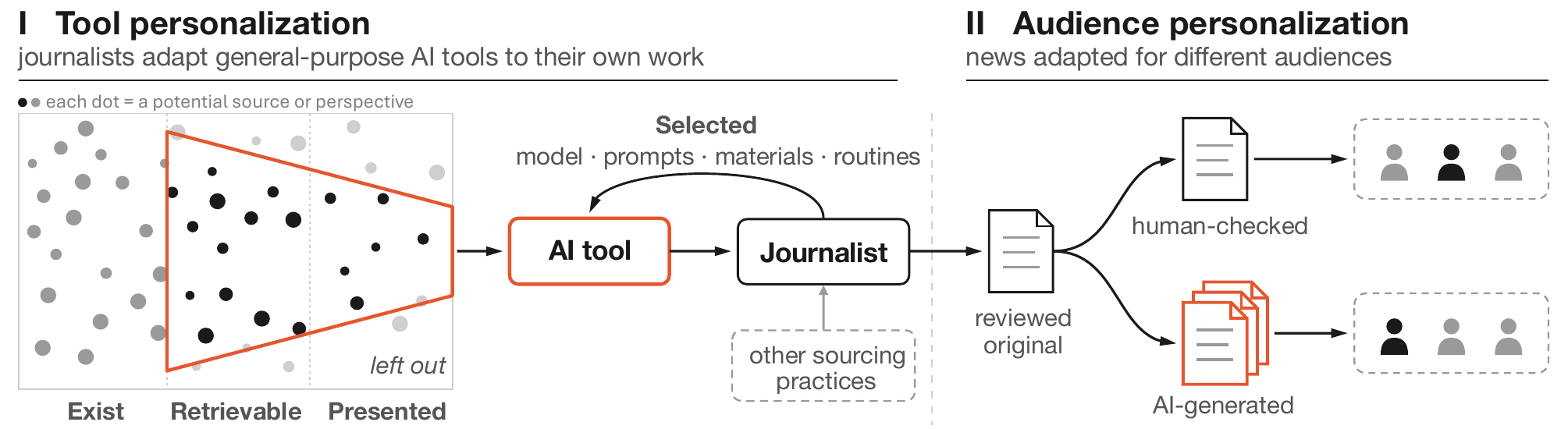}
  \caption{Two-way personalization in AI-assisted news work. (I) Tool personalization: practitioners adapt AI tools through individual configurations and practices, shaping which sources and perspectives become available for editorial judgment. (II) Audience personalization: news is adapted for different readers, with varying degrees of human review.}
  \Description{Two-part diagram. Left: each dot represents a potential information source or perspective. A funnel illustrates which sources an AI-assisted workflow can retrieve and present (black dots) and which remain outside it (grey dots). The AI tool passes information to a journalist, who configures the tool through model choice, prompts, materials, and routines, while also drawing on other sourcing practices. Right: a journalist's reviewed original branches into human-checked and AI-generated versions distributed to different audience groups.}
  \label{fig:twoway}
\end{figure}

\textbf{An empirical account of two-way personalization.}
We show how practitioners adapt news for audiences and AI tools to their work (Figure~\ref{fig:twoway}). They regard changes to news \textit{form} as more readily delegable to GenAI than changes to \textit{meaning}, yet this boundary is difficult to maintain in practice. They also exercise editorial judgment through personally configured tools that may fall outside organizational oversight. Practitioners frame the risks of these two forms of personalization differently: they anticipate diversity risks in audience personalization, while treating limitations in their own AI-assisted sourcing primarily as accuracy problems that better prompting or more capable models could resolve.

\textbf{An extension of news diversity evaluation.} Building on Napoli’s dimensions~\citep{napoli1999}, we identify additional sites where GenAI shapes news diversity: 
(1) in the AI tools newsroom staff use to find and research stories, which shape the sources and perspectives that reach them before a story is written; (2) \emph{within} a single story across AI-generated versions, concerning what essential information is preserved for different readers. 
We further identify \emph{audience categorization} and \emph{distributive fairness} as concerns: how AI systems categorize audiences shapes whose needs are recognized and served, while the distribution of journalistic resources determines which audiences receive human-produced or human-reviewed journalism and which rely primarily on AI-generated content.
Together, these extend diversity analysis beyond the news item as a stable unit to the processes through which journalistic information is selected, adapted, and distributed.

\section{Background}

\subsection{News Personalization and Diversity}
\label{sec:background_diversity}

The tension between news personalization and diversity predates GenAI. As news organizations tailored their offerings to individual preferences, concerns arose about selective exposure, filter bubbles, and the fragmentation of shared public knowledge~\citep{thurman2012,zuiderveen2016}. Research has examined this tension primarily through news recommendation, treating exposure diversity as a design objective~\citep{helberger2018,helberger2019,burgess2025diversity}, developing metrics for evaluating recommended items~\citep{vrijenhoek2021}, and comparing algorithmic with editorial selection~\citep{moller2018}. Organizational studies show how editorial values and institutional priorities shape recommender design~\citep{bodo2019,bastian2021,schjotthansen2023}, including tensions between personalized services and public service commitments to universality~\citep{vandenbulck2018,vrijenhoek2024diversity}. Across this literature, the central question is which existing news items reach which audiences.

The information audiences encounter is, however, shaped well before distribution. Journalists' sourcing practices determine which voices and perspectives enter reporting: digital technologies reshape source selection and verification~\citep{lecheler2016}, professional routines and access constraints limit the sources consulted~\citep{hertzum2022}, and newsrooms have developed ways to monitor source diversity~\citep{grannan2025}. Adaptation also shapes information: structured journalism assembles different news products from shared units of reporting~\citep{caswell2019}, raising questions about preserving meaning when journalistic information is recombined~\citep{jones2019}. These studies locate consequential information choices throughout production, beyond the selection and distribution of finished news items.

\subsection{Generative AI in News
Production and Personalization}
\label{sec:background_genai}

GenAI is entering news production from information gathering and research to drafting and editing~\citep{cools2026uses,diakopoulos2024newswork,cifliku2026local,wasdahl2026algorithms}. 
Its integration depends not only on organizational deployment but also on how practitioners incorporate it into their work. Individual experimentation can develop outside established workflows and prove difficult to coordinate~\citep{xiao2025, xiao2026}; knowledge silos and mismatches between institutional tools and professional needs further complicate adoption~\citep{dodds2025,ali2026adoption,cools2026transformation,lehner2025change,moller2025roles}.
Research on AI assistants also finds discrepancies between automated evaluations and users' preferences for adapted responses~\citep{dong2026memory,zhang2026recentering}, highlighting the importance of evaluating how people assess and use these tools. FAccT scholarship further connects AI integration to questions of editorial authority, examining tensions around ownership, data sharing, and local control~\citep{tseng2025,pedersen2026trust}, as well as the redistribution of editorial decision-making and responsibility~\citep{sorrentino2026}.

GenAI extends personalization beyond recommending existing stories to shaping both the information journalists encounter and the versions audiences receive. Unlike search engines that rank sources~\citep{trielli2019search}, GenAI synthesizes them, making omissions harder to detect~\citep{shah2022situating}. For audiences, generative recommenders combine articles into personalized narratives~\citep{gao2024generative}, while language models adapt reports for different readers. Research has examined political slant in rewriting~\citep{cheng2025balancing}, oversight of generated versions~\citep{jay2025}, and audience interest in summaries and translations~\citep{rossarguedas2025}. 
Audience acceptance depends on transparency, human oversight, and where AI enters production~\citep{thomas2026expecting,thomson2025report}. 

\subsection{Editorial Jurisdiction, Oversight, and Accountability}
\label{sec:background_gov}

Abbott defines professional jurisdiction as the link between a profession and its work: the authority to define a task's standards, perform it, and judge its outcomes~\citep{abbott1988}. New technologies can unsettle this jurisdiction by making tasks delegable, prompting professions to engage in \textit{boundary work} over what remains within their authority~\citep{gieryn1983,carlson2015}. This lens directs attention to which decisions journalists retain when AI performs parts of their work, and whether retaining authority in principle translates into control in practice.

Editorial control is also central to emerging AI governance. The EU Digital Services Act requires transparency in recommender systems, systemic risk assessments addressing media pluralism for very large platforms, and a non-profiling recommendation option~\citep[Arts.~27, 34, 38]{dsa2022}. The EU AI Act exempts AI-generated public-interest text from disclosure requirements when it undergoes human review or editorial control and a person holds editorial responsibility~\citep[Art.~50(4)]{aiact2024}. The Council of Europe further calls for editorial oversight of both AI-generated outputs and the processes producing them~\citep{coe2023aijournalism}. 
Governance approaches also differ across regions. China's AI-generated content labeling rules require explicit and implicit identification of synthetic content~\citep{chinaLabeling2025}, while the ASEAN Guide on AI Governance and Ethics offers voluntary recommendations for responsible AI deployment~\citep{asean2025aiGovernance}. 
Formal editorial control, however, differs from meaningful oversight, which requires the capacity to inspect and influence consequential decisions~\citep{sterz2024}, while accountability requires that such decisions can be explained and challenged~\citep{bovens2007,wieringa2020}. Under two-way personalization, decisions shaping news diversity may occur through individually configured tools during sourcing or across multiple audience-specific versions. Formal editorial responsibility for published news therefore does not necessarily ensure effective control over the decisions that shape its content and diversity.


\section{Methods}
\subsection{Formative Roundtable}
We began with a facilitated exploratory roundtable in June 2026, bringing together nine professionals: seven from editorial, technical, product, and strategy roles in German public service and commercial media, and two senior professionals from the Asia-Pacific region.
Discussions covered organizational challenges and technical solutions for AI-driven news personalization. 
With participants' consent, roundtable insights informed the interview protocol but were excluded from empirical analysis; four participants were later interviewed.

\subsection{Interviews and Participants}
We then conducted 16 semi-structured interviews (Table~\ref{tab:participants}) using maximum variation purposive sampling~\citep{palinkas2015purposeful} across professional roles, organizational models, and institutional contexts. Given that journalistic practices are shaped by individual, organizational, and systemic influences~\citep{reese2016media}, with substantial variation even within national media systems~\citep{hovden2021nordic}, we prioritized heterogeneity over numerical balance. Participants held editorial, technical, product, strategy, and regulatory roles in public service, public--private, commercial, and state-affiliated media, technology providers, and cross-sector media organizations in Europe and the Asia-Pacific region. These settings encompass different conditions for news production and distribution, including public service mandates, commercial pressures, and platform-mediated distribution. Recruitment also drew on the authors' professional networks and linguistic and cultural familiarity~\citep{robinson2014sampling}. Rather than seeking representative comparisons between groups, the design allowed us to examine recurring judgments alongside context-specific differences, interpreting participants' accounts in relation to their professional and organizational settings.

Interviews lasted 60--90 minutes and explored professional responsibilities, success metrics, editorial values, content adaptation, AI use, and news diversity, with probes adapted to participants' roles and responses (Appendix~\ref{app:interview-guide}). A hypothetical AI-assisted personalization tool prompted discussion of editorial control, accountability, human oversight, and content traceability. Four interviews were conducted in Chinese, one in German, and the remainder in English. Non-English transcripts were prepared by native speakers and cross-checked by a second native-speaking team member.
The study received joint approval from the two institutional research ethics committees where team members were based. Participation was voluntary and uncompensated. Interviews were recorded with informed consent, transcribed, and pseudonymized.

\begin{table}[t]
\centering
\caption{Participant characteristics.}
\label{tab:participants}
\small
\begin{tabular}{@{}lllll@{}}
\toprule
ID & Professional role & Organization type & Region & Experience \\
\midrule
P01 & Technical               & Media technology / services              & Europe, Germany        & $\leq$5 years \\
P02 & Management              & Public--private broadcaster       & Europe, Germany        & $>$15 years \\
P03 & Strategy \& Technical   & Media technology / services          & Europe, Germany        & 6--15 years \\
P04 & Editorial               & Commercial news organization                   & Europe, Germany        & 6--15 years \\
P05 & Production              & Public broadcaster                & Europe, Germany        & $>$15 years \\
P06 & Production              & Public broadcaster                & Europe, Germany        & $>$15 years \\
P07 & Editorial \& Management & Public broadcaster                & Europe, Germany        & 6--15 years \\
P08 & Editorial               & Public broadcaster                & Europe, Germany        & $\leq$5 years \\
P09 & Editorial               & Public broadcaster                & Europe, Germany        & $\leq$5 years \\
P10 & Editorial \& Technical  & Public broadcaster                & Europe, Germany        & $\leq$5 years \\
P11 & Editorial               & Commercial news organization                   & Asia-Pacific, China    & $\leq$5 years \\
P12  & Editorial               & State-affiliated news organization             & Asia-Pacific, China    & $\leq$5 years \\
P13 & Editorial               & Commercial news organization                   & Asia-Pacific, China    & $\leq$5 years \\
P14 & Editorial \& Strategy   & Media technology / service  & Asia-Pacific, China    & 6--15 years \\
P15      & Legal \& Regulatory     & Cross-sector media organization   & Asia-Pacific, Malaysia & 6--15 years \\
P16       & Management              & Cross-sector media organization   & Asia-Pacific, Malaysia & $>$15 years \\
\bottomrule
\end{tabular}
\end{table}

\subsection{Analysis}
\label{sec:analysis}

We prioritized participants' professional practices and experiences while distinguishing them from broader beliefs and expectations~\citep{sundler2019qualitative}. 
Two researchers independently open-coded an initial transcript and compared codes and analytical memos; one then coded the full dataset and the other independently coded ten transcripts. Both reviewed each other's coding and refined codes through discussion rather than inter-rater reliability metrics~\citep{mcdonald2019reliability}.
Each of 940 excerpts was classified as included (relevant to the research questions, 537), context (background needed to interpret them, 351) or excluded (52). 
A researcher used Claude Opus 5.5 to identify semantic overlaps and consolidate the initial codes into 141 codes across three iterative rounds, refining the codes at each stage to reduce redundancy while preserving meaningful distinctions. The researcher reviewed 72 flagged decisions and revised eight (see \hyperref[sec:genai-usage]{Generative AI Usage Statement}).
The full research team then organized the codes into 38 clusters and developed themes collaboratively~\citep{braun2006using,braun2021one}.
The analysis combined inductive coding with abductive theorizing~\citep{timmermans2012abductive}: personalization, diversity, editorial control, and accountability served as sensitizing concepts~\citep{blumer1954,bowen2006sensitizing}, and as patterns emerged we drew on professional jurisdiction and boundary work~\citep{abbott1988,gieryn1983} to explain which decisions participants claimed or delegated, and on Napoli's dimensions~\citep{napoli1999} to locate where diversity-relevant decisions arose.
We retained disconfirming accounts and tagged excerpts as practices, experiences, observations, concerns, expectations, or wishes.

\subsection{Validation}
To complement participants' self-reported accounts, we conducted targeted validation studies to independently examine selected claims about AI-assisted news production and personalization. These studies assess observable model behaviour under controlled conditions, providing additional empirical evidence to contextualise the interview findings. Detailed procedures and results are reported in Appendix~\ref{app:validation}.

\section{Findings}
Our findings trace how GenAI reshapes journalistic decision-making in ways that existing approaches to evaluating and governing news diversity fail to capture. Participants defended professional jurisdiction over editorial meaning while accepting the delegation of form in \textit{audience personalization} (Section~\ref{sec:professional_jurisdiction}). They also exercised this jurisdiction through \textit{personalized AI tools}, within the broader organizational and industry ecosystem (Section~\ref{sec:tool_personalization}). The two forms of personalization revealed different understandings of risk: risks to audiences were largely anticipated, while diversity-relevant problems encountered with personal AI tools were understood primarily as issues of accuracy and capability (Section~\ref{sec:experienced_anticipated}). Together, these findings bring into view diversity-relevant decisions that conventional evaluations of published news cannot capture (Section~\ref{sec:diversity_relevant_decisions}). They also expose a governance challenge: journalists remain responsible for outcomes shaped by decisions they may not fully observe or control (Section~\ref{sec:accountability}).

\subsection{Audience Personalization as Professional Jurisdiction}
\label{sec:professional_jurisdiction}

\textbf{Audience adaptation as established professional work.}
Understanding and responding to audiences is integral to journalism~\citep{kovach2014,costerameijer2020}. Participants described adaptation as everyday editorial judgment over language, examples, emphasis, and the social context in which news is received. As P09, a journalist working across different audiences, explained:

\begin{quote}\small
It's actually one of our main tasks to always adapt to the target outlet or the target audience. [\ldots] You're trying to adapt it to the language style, but also, like, again, relevance. So for a young audience, I would try to find an example out of the daily lives of young people and vice versa for older people. (P09)
\end{quote}

\noindent She illustrated this with research on music and recovery from surgery, framed through Taylor Swift for younger listeners and Metallica for older ones. Adaptation also required sensitivity to social contexts. P11 emphasized ``taking into account the emotions of your readers, also certain sentiments in the contemporary society.'' Such judgment was also part of professional identity. Reflecting on a tool that could personalize her articles, P08 acknowledged its potential to improve understanding, but added, ``even if it would be perfect, it would be a little bit sad for me because I like adapting.'' GenAI thus enters an established domain of editorial expertise that practitioners both exercise and value.

\noindent\textbf{Jurisdiction shapes what counts as a diversity concern.}
Participants associated audience personalization with echo chambers and the loss of shared information, but their views on how to address these risks reflected who controlled news distribution. 
At public and public--private broadcasters with their own distribution services, participants (P02, P05, P07, P08, P09) discussed ways to improve exposure diversity through recommendation design: P02 suggested a wildcard so that among five recommended items ``one is out of my echo chamber'', while P05 wanted access to ``the full menu, not only the user-for-me generated content''.

Where news distribution depended on publishers or third-party platforms, participants placed responsibility for recommendation elsewhere.
P04, a commercial-news editor, described recommendation as ``not my business. It's distribution and our publisher has to decide.'' 
P13 described the same boundary in Chinese platform-mediated distribution:
 
\begin{quote}\small
Suppose our organization uploads ten videos about a landslide in Nepal. It is not the news organization that decides who sees which video; the platform does, through its recommendation algorithm. That question is better directed to Douyin, Xiaohongshu, or WeChat. (P13)
\end{quote}

\noindent P13 instead emphasized diversity within reporting, describing strong journalism as examining a subject ``in depth and from several angles''. 
These accounts suggest that professional jurisdiction shapes which dimensions of diversity practitioners foreground: those with influence over recommendations discussed broadening exposure, while others located responsibility for distribution elsewhere and emphasized diversity within reporting.

\noindent\textbf{Delegating form, retaining meaning.}
Participants distinguished three areas of editorial decision-making in GenAI personalization. \emph{Form}, including language, length, format, and tone, was considered delegable as long as the underlying information remained intact. P03 accepted adaptation ``to reach more people for greater good'', but rejected ``creating more stuff''. \emph{Selection}, concerning which stories reach which audiences, was more contested, reflecting authority distributed among journalists, editors, publishers, and platforms. \emph{Meaning}, including factual claims, interpretation, and stance, remained a core editorial responsibility. P04 resisted adaptation that could produce ``a complete turnaround of the arguments and maybe also of the facts'', while P02 insisted that an original interview ``has to be the baseline for truth''.

These distinctions constitute professional \emph{boundary work}~\citep{gieryn1983}: practitioners define which decisions AI may perform and which remain under editorial authority. Delegating execution need not surrender professional jurisdiction if journalists retain the authority to set standards and evaluate outcomes~\citep{abbott1988}. Similar boundaries appear in audience research, where AI-assisted editing, translation, and accessibility are more acceptable than AI involvement in authorship and editorial judgment~\citep{thomas2026expecting,thomson2025report}.

However, the boundary between form and meaning is difficult to maintain in practice. P11 described how shortening a report for social media could omit important context or introduce the adapter's interpretation. P05 illustrated how even a single word could change what a report communicates:

\begin{quote}\small
If you say, for example, they opened their shops again, it is just neutral. If you say they could open their shops again, you kind of draw a little screw in the listener's mind, which says, oh, they were attacked, you know? (P05)
\end{quote}

\noindent Meaning is partly shaped by the very presentational choices practitioners consider delegable, and GenAI enables adaptation at a scale that may exceed editorial review. As P02 anticipated, with ``a thousand versions or a thousand users'', journalists cannot ``click and look and read every little bit which is in there.'' Professional jurisdiction therefore depends not only on which decisions journalists claim as their own, but also on whether they can exercise that authority when AI performs the work.

\subsection{Tool Personalization as Editorial Practice}
\label{sec:tool_personalization}

\textbf{Personalization moves upstream into news production.}
Practitioners adapted general-purpose AI tools to research and develop stories, shaping what information and perspectives entered their work. P11 described delegating background research, previously conducted through keyword searches, to AI:

\begin{quote}\small
Now, I've delegated this task to AI tools like ChatGPT and Claude to help me organize the information first. However, I wouldn't say I use all of their organized content directly, because they often make mistakes [\ldots] So basically, after I have it gather information for me in the early stages, its role is to help me form a preliminary understanding of the story, so I can get a rough idea of the potential twists and angles that might not have been covered yet. (P11)
\end{quote}

\noindent AI thus became involved in the early stages of reporting, while practitioners actively configured the tools. P13 instructed models ``to think beyond the structure I have proposed'', recognizing that ``it tends to agree with your initial approach''. P09 described learning to work effectively with AI through repeated use: ``co-working with the AI is, like, not something natural''.
We conceptualize these practices as \textbf{tool personalization}: practitioners adapt general-purpose AI systems through model choice, source materials, prompting, and recurring routines to suit their professional needs and judgments. Through these adaptations, AI becomes part of how practitioners gather information and develop potential story angles.

\noindent\textbf{Tool personalization as an exercise of professional jurisdiction.}
These adaptations were guided by practitioners' existing professional knowledge and standards. Rather than accepting generic outputs, they decided what context to provide, how to direct the model, and what constituted an acceptable result. P07 illustrated this when describing how she prepared AI-assisted responses to audience complaints:

\begin{quote}\small
I say, ``So-and-so, the colleague who wrote the article about the Hungarian power station, is an all-rounder, a correspondent, not necessarily an expert in reactor safety, so please reply like this, like this, like this; take this and that content into account.'' I hand over this package [\ldots] and then have it generate a draft reply, which I then double-check and consider: ``Yes, is that spot on, or does it miss the mark slightly?'' And I also tweak it again, for instance, simply removing the typical AI-style elements. (P07)
\end{quote}

\noindent Here, personalization begins with professional judgment: P07 supplies knowledge the model lacks, specifies an appropriate response, and evaluates the draft against her editorial standards. Similar judgments shaped how participants adapted AI-generated language. P14 revised expressions that sounded ``too official, inflated, or unnatural'', while P09 warned that recognizable AI prose was ``cutting your own flesh'' because it could alienate readers. These practices show how practitioners bring established expectations about accuracy, appropriateness, and authorship into their use of AI. 
Tool personalization thus becomes a means of exercising professional jurisdiction: practitioners retain authority over how AI is instructed, what standards its outputs must meet, and which contributions enter their work. The same configurations also shape the information on which subsequent editorial judgments depend. Configuring tools around familiar assumptions may reinforce those assumptions, even when practitioners intend to improve their work.

\noindent\textbf{Individual adaptation alongside organizational systems.}
The ability to personalize AI tools also depended on organizational arrangements. Tool personalization persisted even where organizations provided internal AI infrastructure. P08 described a public broadcaster's internal toolbox designed so that ``we don't send our data to the US'', allowing staff to use AI ``more relaxed'', although such sanctioned tools did not always meet individual professional needs. P07 explained why she continued using her own AI assistant:

\begin{quote}\small
My employer has restrictions on what I'm allowed to use and what I'm not. That's why I might use my own AI in the workplace. We have an AI editor, which is very good, but it doesn't cover all our current needs. [\ldots] Development is so fast-paced at the moment that, as a public broadcaster, we naturally can't always keep up, and in some cases, we're missing out on features or simply the personalization you get with your own AI tailored to your needs. (P07)
\end{quote}

\noindent Individual adaptation thus reflected a gap between standardized institutional tools and evolving professional needs, not simply resistance to organizational systems. 
Accessible general-purpose models made such experimentation possible, while informal organizational expectations could encourage it: P12 recalled a manager suggesting, ``If you can't think of anything, ask AI and give me a dozen questions.''

These arrangements enabled practitioners to develop tools around their professional judgments, even when organizational provision lagged behind their needs, but the resulting configurations and routines could remain outside shared organizational review. Tool personalization thus supported individual editorial autonomy while potentially limiting organizational visibility into the decisions shaping journalistic work.

\subsection{How Practitioners Understand Risks Across Two-Way Personalization}
\label{sec:experienced_anticipated}
Participants framed risks differently across the two-way personalization. Diversity limitations encountered in their own AI tools were predominantly treated as accuracy and capability problems, whereas audience personalization prompted concerns about news diversity and shared public knowledge. 
These different framings reflect the boundaries of professional jurisdiction: practitioners approached their own AI-assisted workflows through standards of professional competence, while evaluating audience personalization against journalism's broader public responsibilities.

\noindent\textbf{Experienced tool limitations as problems of professional craft.}
Nine participants (P05, P07--P09, P11--P15) described limitations encountered in research and production, including fabricated information, incomplete retrieval, generic responses, and summaries that omitted important context. Some limitations concerned which information could enter reporting. P11 described difficulties accessing non-dominant-language sources and content on platforms poorly accessible to AI systems:

\begin{quote}\small
The first type of information it can't access is content in languages other than Chinese and English. [\ldots] I asked [the model] to help me find something in Arabic because I was looking for information on a foundation in the Middle East, but it seems my AI can't find [in Arabic]. [\ldots] Another type of information it can't find is content from [\ldots] WeChat Official Accounts. [\ldots] A lot of content in China, including corporate announcements and news from new media outlets, is published exclusively on Official Accounts, so I have to find that information myself. (P11)
\end{quote}

\noindent Other limitations concerned the sources and perspectives models presented. P13 observed that models tended to agree with the user's initial approach; P14 found that an apparently useful source could be ``a very obscure outlet merely reposting someone else's material''; and P11 reported that, when researching a topic she had previously covered, ``the AI's feedback consisted entirely of my own previous coverage''. These limitations concerned not only factual correctness but also the breadth of information available for journalistic judgment.

Participants nevertheless approached these problems primarily through professional craft. Their responses involved checking original materials, providing context, revising prompts, or completing the task themselves. When an AI draft became absorbed in irrelevant technical details, P07 concluded, ``That's obviously a prompt error''. After repeatedly explaining context, P11 decided, ``I just go ahead and write it myself''. Within this framing, the problem was an inadequate instrument, and the remedy was further professional intervention. Tool personalization restored control rather than prompting evaluation of its broader informational consequences. Photo editors, by contrast, have framed generative image risks in normative terms such as stereotyping and bias~\citep{thomson2025visual}. The difference is one of visibility: representational bias appears in a generated image, whereas an omission from AI-assisted research leaves no trace in the output.

\noindent\textbf{Anticipated audience harms as problems of public responsibility.}
Audience-facing personalization, by contrast, prompted concerns about news diversity, echo chambers, and shared public knowledge. P01 worried that when ``everyone gets different news [their] image of the world can then change'', while P04 defended ``the things that everyone has to know''. P07 articulated the public service concern:

\begin{quote}\small
Individual AI models will soon be churning this out, or are, of course, already doing so. But I'm not sure if that's the recipe for public service journalism, because I consider it very important that there's still a shared space. That `campfire' aspect, where people still chat about shared experiences of information. (P07)
\end{quote}

\noindent These concerns reflected practitioners' normative commitments to journalism's public role, rather than observed effects of generative personalization on their audiences. Practitioners could articulate what journalism ought to provide, including shared knowledge and diverse perspectives, without necessarily having sufficient audience evidence to establish how personalization would affect these outcomes. P10 described the tension between needing to understand audiences and avoiding intrusive data collection: ``we, of course, need to know who are we talking to and who we're making our content for'', while ``we don't want to scare off our users'' or appear as ``data sharks''. P16 noted that many broadcasters relied on social media data because they could not afford audience measurement, while questioning whether such data were representative.

Assumptions about audience responses to AI also differed. P07 believed that ``if people realise it's an AI-generated text, they stop reading'', while P06 expected that ``the tolerance of people to accept an AI story might just increase''. Following Nelson~\cite{nelson2021}, we understand these as imagined audiences: working assumptions about public expectations that inform editorial decisions in the absence of direct evidence. Audience research both supports and complicates these assumptions: audiences express strong expectations for disclosure and may interpret undisclosed AI use as deception~\citep{thomson2025report,thomas2026expecting}, although disclosure itself can reduce perceived credibility~\citep{thomas2026expecting}.

This difference in risk framing was predominant among participants (detailed in Appendix Table~\ref{tab:matrix}). Fifteen participants discussed both forms of personalization, though not always from direct experience. Even when participants recognized bias or restricted perspectives in their own tools, they generally framed these as problems of reliability or professional craft rather than news diversity. Meanwhile, some located responsibility for audience personalization with platforms or clients rather than journalists. Practitioners thus encountered concrete tool limitations they could attempt to remedy, while articulating broader diversity commitments for audiences whose responses to generative personalization remained largely untested. Professional jurisdiction shaped not only how practitioners used AI, but also which risks they understood as matters of professional competence and which as responsibilities to the public.

\subsection{Where Diversity-Relevant Decisions Arise under GenAI}
\label{sec:diversity_relevant_decisions}

\begin{figure}[ht]
  \centering
  \includegraphics[width=\linewidth]{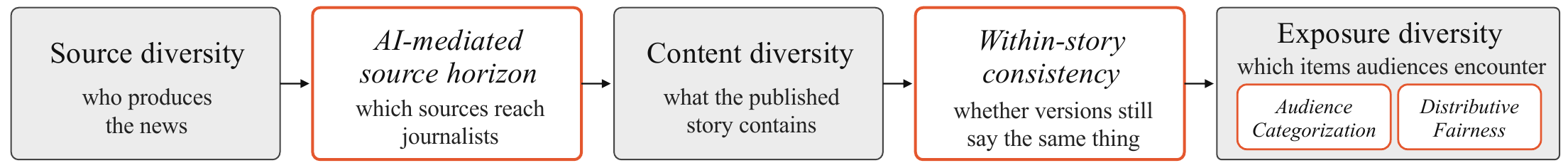}
  \caption{Extending news diversity under GenAI. Grey: Napoli's dimensions~\citep{napoli1999}; orange: extensions identified in this study.}
  \Description{A left-to-right chain of five boxes: source diversity, AI-mediated source horizon, content diversity, within-story consistency, and exposure diversity, with exposure diversity containing two sub-boxes for audience categorization and distributive fairness. The new concepts are highlighted in orange.}
  \label{fig:dimensions}
\end{figure}

Our interviews suggest that decisions shaping news diversity are not always recognized as such, nor are their consequences necessarily visible in published news. Two phenomena emerged: limitations in AI-assisted sourcing are often framed as matters of accuracy or professional craft, despite shaping which perspectives enter reporting; meanwhile, generative adaptation may subtly alter what different versions of the same story convey. We further examined these mechanisms through two targeted validation studies (Appendix~\ref{app:validation}) and considered their implications for where diversity is shaped, how it is evaluated, and which decisions remain open to editorial scrutiny. Figure~\ref{fig:dimensions} extends Napoli's framework to locate these additional sites.

\noindent\textbf{AI-mediated source horizons.}
AI-assisted research shapes which sources and perspectives reach editorial consideration, a process situated between Napoli's source and content diversity~\citep{napoli1999}. We refer to this range, shaped by retrieval capabilities, model behavior, and journalistic practice, as the \emph{AI-mediated source horizon}. From our interview analysis, we identify four distinct layers: (1) the \emph{sources that exist}, including original interviews and local knowledge unavailable online, on which P13 noted exclusive reporting depends; (2) the \emph{sources a system can retrieve}, constrained by language coverage, platform access, and available content, with P10 noting that retrieval systems' content stores were often not ``elaborate enough'' for independent use; (3) the \emph{sources and perspectives a system presents}, shaped by prompting and model behavior, as P09 observed when differently framed questions produced opposing answers supported by sources from the same outlet; and (4) the \emph{sources the practitioner selects and uses}, with P12 prioritizing government websites, major news organizations, and academic journals. Practitioners can expand this horizon by supplying documents, reformulating questions, or reporting independently.

We observed that AI-assisted research can broaden sourcing beyond established journalistic routines. P07 noted that journalists can fall into ``routines that aren't actually that good from the point of view of diversity of perspective'', particularly when repeatedly consulting familiar experts. Nonetheless, AI-assisted and conventional sourcing should not be treated as fixed alternatives: as AI becomes embedded in journalistic work, it may reshape the very routines against which its effects on diversity are evaluated.
Compared with search engines and other conventional sourcing practices, GenAI mediates not only which sources journalists encounter but also how information is extracted and synthesized, making exclusions harder to observe (Appendix~\ref{app:search-contrast}).
Published-source counts cannot reveal what journalists never encountered, just as factual verification cannot establish whether relevant perspectives were considered.

Our illustrative verification cases demonstrate how information can disappear at different layers (Appendix~\ref{app:horizon}). In one case, a publicly accessible German investigation was retrievable but not surfaced in response to an open question: two of three systems failed to locate the relevant figures, while the third supplied incorrect ones. Providing the source URL enabled all three to answer correctly, demonstrating both the value and the limits of supplying known sources. Other cases showed platform access barriers, substitution of unsupported information, and omitted supplementary findings. These cases illustrate mechanisms of exclusion rather than their prevalence, and help explain why missing sources may remain invisible even when AI-assisted answers appear complete and accurate.

\noindent\textbf{Within-story informational consistency.}
News diversity evaluations typically examine which news items audiences encounter~\citep{helberger2018,vrijenhoek2021}, treating each item as a relatively stable unit of information. Generative adaptation complicates this assumption: different audiences may encounter versions of the same story that retain its main idea but differ in essential information. In practice, concerns about such adaptation were also framed primarily in terms of accuracy rather than diversity. P14 described producing ``ten short versions'' from a master article and checking them ``for factual accuracy.'' Yet factual accuracy alone does not establish whether each version preserves the information essential to understanding the original report. We examine this as \emph{within-story informational consistency}: the preservation of core claims, evidential qualifications, and context across versions, rather than uniformity of presentation.

Our adaptation study examined informational consistency both between each adapted version and its original report and across versions generated for different audiences (Appendix~\ref{app:consistency}). We adapted 12 articles from Europe and the Asia-Pacific for five audience types, comparing a basic adaptation instruction with a variant explicitly requiring meaning preservation. An LLM evaluator assessed whether adaptations changed the main idea, with human validation. We separately assessed retention of 2--4 essential information units per article, defined as information necessary to understand its central claims, evidential qualifications, and context. These units were identified by a researcher and cross-checked by a journalist. A manual review of 103 retention judgments showed substantial agreement with the automated evaluator (Cohen's $\kappa=0.821$).

The evaluations showed that LLMs generally preserved the main idea of each article, with no omissions judged to alter it. However, at the finer level of essential information, versions retained approximately 75\% of designated units across models, while explicit preservation instructions produced limited improvements (approximately 5 percentage points on average). Across audience versions, the information retained also differed. For example, under Gemini, versions for B1-level readers and specialists had a mean Jaccard distance of 0.323, indicating that, on average, approximately one-third of the distinct essential units retained across each pair of versions were not shared by both. Thus, versions preserving the same main idea could still provide audiences with different subsets of essential information.

These findings expose a limitation of item-level evaluation: two readers counted as exposed to the same story may receive materially different information. The relevant unit of evaluation therefore extends from individual news items to the relationship between a reviewed report and its generated versions.

\noindent\textbf{Further concerns: audience categorization and distributive fairness.}
The interviews raise two further concerns for diversity evaluation. The first concerns \emph{audience categorization}: personalization presupposes decisions about whose needs are recognized and how. P09 envisioned personalized news stories for households where German is not the first language. Regarding how audience preferences are obtained, P10 favored asking users directly rather than inferring their preferences. Classification systems embed choices about who becomes visible~\citep{bowker1999}; evaluating exposure diversity within predefined groups therefore overlooks how those groups themselves shape what audiences receive.

The second concerns \emph{distributive fairness}: what kinds of journalism different audiences receive. Participants described or anticipated automation in routine, regional, and off-peak services, including overnight news (P05), weather (P09), and minority-language bulletins (P16). Others emphasized retaining human involvement in flagship news (P02) or services where audiences value human companionship (P08). Automation can expand access where human-produced services are limited~\citep{thomas2026expecting}, but may also distribute editorial attention and verification unevenly. If certain audiences receive primarily unreviewed generated content, they may bear a disproportionate share of its informational risks, echoing concerns about the unequal distribution of automation's consequences~\citep{eubanks2018}.

\subsection{Divided Control and the Limits of Accountability}
\label{sec:accountability}

The professional jurisdiction defended by participants carried obligations of accountability: retaining authority over editorial meaning meant accepting responsibility for published content. Six participants (P01, P02, P07--P10), mostly responding to hypothetical scenarios, assigned responsibility for AI-related errors to a human author or editor. P08 illustrated this professional expectation through a case involving an AI-generated image in a news report:

\begin{quote}\small
She was responsible for that so she had to leave her position [\ldots] I do feel pity for because obviously she was very busy and stressed out and that was a mistake but it's her responsibility and that's what we do as journalists, we own our stories. (P08)
\end{quote}

\noindent The capacity to fulfil this obligation, however, was unevenly distributed. P12 described the limits of verification for junior staff: ``With my pay and resources I can only verify so far. [\ldots] What I can demonstrate is that every sentence in my article has a source.'' These accounts expose a tension between professional responsibility and the practical capacity to exercise control, particularly among front-line staff with limited time and resources for verification.

Organizational responsibility is conditional on procedural compliance. P07 located responsibility with management, arguing that organizations cannot ``put 15,000 people in place to double-check the texts generated by the AI'', but qualified this position when employees used AI negligently or failed to follow approval procedures. This presumes that organizational procedures can adequately govern actual practice. As Section~\ref{sec:tool_personalization} shows, limited institutional capabilities, unmet professional needs, and informal expectations sustain AI use beyond standardized workflows. Responsibility can therefore return to individuals precisely where organizational arrangements make compliance difficult to sustain or verify. This resembles Elish's \emph{moral crumple zone}~\citep{elish2019}: front-line practitioners may bear responsibility for outcomes shaped by systems and conditions they do not fully control.

Diversity-relevant decisions, moreover, rarely surface as identifiable errors. AI-mediated sourcing shapes which perspectives enter editorial consideration, while generative adaptation alters what different audiences receive. Such omissions may escape scrutiny even when individual outputs are accurate and approved. Moreover, expanding human review without preserving the expertise and conditions required for meaningful judgment risks reproducing Bainbridge's \emph{ironies of automation}~\citep{bainbridge1983}. The accountability problem is therefore not who approves outputs, but whether those held responsible can recognize, scrutinize, and influence the decisions that shaped them.

\section{Discussion}

Our findings show that GenAI relocates personalization, and with it the decisions that shape news diversity, to sites where they are harder to observe and where formal responsibility and practical control diverge. Diversity-relevant decisions now arise in the tools through which journalists research and in the versions through which stories reach different audiences, ranging from sources never encountered to variations too numerous to review. Practitioners claim jurisdiction over editorial meaning, yet increasingly exercise judgment through individually adapted tools and workflows. The challenge is therefore not simply whether journalists retain editorial authority, but whether the decisions shaping what they can know and communicate remain visible and contestable.

\subsection{Evaluating News Diversity Under GenAI}
Existing approaches evaluate news diversity through published content and its distribution, including source and viewpoint counts or diversity metrics for recommended items~\citep{vrijenhoek2021}. Our framework extends evaluation to the information available during production and the relationships among versions of a story.

For AI-assisted research, this means comparing sources and perspectives available through conventional journalistic practices, general-purpose models, and individually configured tools on comparable tasks. Such comparisons should assess linguistic and platform coverage, source independence, local knowledge, and the effects of prompting and retrieval. Evaluation must distinguish what systems make available from what practitioners select, and capture expansion as well as exclusion: AI may introduce perspectives journalists would otherwise overlook while obscuring others.

For audience-facing adaptation, evaluation must examine the relationship between a reviewed report and its generated versions. The criterion is not identical wording but preservation of core claims, evidential qualifications, and context. Our adaptation study suggests one operationalization: measuring retention of essential information separately from whether the main idea survives, since these can diverge. Cross-version comparisons can reveal omissions and shifts in certainty that item-level exposure metrics miss.

Evaluation must also examine whose needs audience categories recognize and what journalism different audiences receive. Automated regional, minority-language, or off-peak services may expand access without equivalent human reporting or verification. Accessibility, informational consistency, and the equitable distribution of editorial attention therefore require distinct assessment.

\subsection{Governing AI-Mediated News Production}

Conventional oversight relies on checking presented information, approving published content, and assigning responsibility to individuals. Our findings reveal the limits of each under two-way personalization.

First, \emph{checking presented information cannot reveal relevant sources that were never retrieved}. Practitioners' verification practices address what AI returns, not necessarily what it excludes. Their remedies also warrant scrutiny: customization can supply missing context while reinforcing familiar sources, framings, and previous coverage. Recurring prompts, assistants, and retrieval configurations should therefore become reviewable organizational practices, assessed for the information they surface or exclude. Sanctioned tools must also meet professional needs, and contextualization and verification must be recognized as work; otherwise, restrictions risk driving consequential practices further out of view.

Second, \emph{approving an original report cannot guarantee informational consistency across generated versions}. Our adaptation study shows that versions preserving the main idea can still omit essential information, and explicit instructions to preserve meaning produced little improvement. Traceability to a reviewed original and checks targeting essential information are therefore necessary when individual review of ``a thousand versions'' is infeasible. Disclosure identifies AI involvement but does not establish substantive preservation. The AI Act's editorial-review exception~\citep[Art.~50(4)]{aiact2024} makes this distinction consequential: meaningful editorial control must extend beyond approving an original report to ensuring that essential information survives adaptation.

Third, \emph{assigning responsibility to journalists does not establish control over the decisions shaping their work}. While participants maintained that journalists ``own our stories'' (P08), they also described limited verification capacity and organizational responsibility conditional on procedural compliance. Effective oversight requires the ability to inspect and influence consequential decisions~\citep{sterz2024}; otherwise, responsibility concentrated on front-line staff risks creating a \emph{moral crumple zone}~\citep{elish2019}. Accountability must extend to those who require AI use, configure systems, and set review conditions. Organizations must resource verification and preserve the expertise on which oversight depends, rather than assume it will survive efficiency-driven adoption~\citep{matich2025oldthreats}. Providers and platforms must likewise make consequential retrieval, ranking, and distribution decisions open to scrutiny.

\section{Limitations}
This study identifies mechanisms rather than the prevalence of practices or their effects across newsrooms. 
Our purposive sample captures diverse professional perspectives, with firsthand accounts of tool personalization but largely anticipatory concerns about audience-facing personalization. 
The evidence matrix (Table~\ref{tab:matrix}) summarizes interpretive coding rather than frequencies of behavior.
Our validation studies illustrate specific mechanisms through selected research tasks, AI systems, articles, and audience adaptations. 
Without a conventional search baseline, our source-horizon cases demonstrate how GenAI shifts exclusion to less observable stages, not whether it excludes more sources than search engines.
Essential-information retention also depends on researcher judgments and automated evaluation. Establishing how frequently these mechanisms affect published journalism, and with what consequences for audiences, requires further studies of real-world newsroom practices.

\section{Conclusion}
Generative AI reshapes news diversity not only through what audiences receive, but also through the information newsroom practitioners encounter and the decisions they delegate. Drawing on 16 interviews, we show that participants seek to retain editorial authority while increasingly personalizing AI tools in ways that make consequential decisions less visible to organizational oversight. Read through established diversity dimensions, these shifts locate diversity-relevant decisions in the source horizons of AI-assisted research and in the informational consistency of personalized versions, before publication and beyond the reach of item-level review. These findings reveal a central tension: front-line staff may remain responsible for information they cannot fully control or scrutinize. Safeguarding news diversity therefore requires moving beyond human approval of published outputs towards accountability for the processes that determine which perspectives enter journalism, how information is transformed, and what different audiences come to know.

\bibliographystyle{ACM-Reference-Format}
\bibliography{references}

\clearpage
\appendix
\section{Interview Topic Guide}
\label{app:interview-guide}

The interviews followed a conversational, topic-guided format. Questions and follow-up probes were adapted to participants' roles and responses, with the order and depth of discussion varying across interviews. The semi-structured guide below was used in our interviews as a rough conversation guide to aid exploration of participants' experiences, reasoning, and perspectives.

\subsection{Professional Practices and Editorial Judgment}

\paragraph{Professional responsibilities and workflows.}
\begin{itemize}
    \item Could you describe your professional role and day-to-day work?
    \item What tools and platforms do you typically work with?
    \item How are decisions typically made in your area of work?
\end{itemize}

\paragraph{Editorial priorities and success.}
\begin{itemize}
    \item What matters most in the work you do?
    \item How do you know when something has worked well?
    \item How does your organization think about success?
    \item Can you think of a time when different priorities came into conflict? What happened?
\end{itemize}

\paragraph{News production and adaptation.}
\begin{itemize}
    \item Does your work involve adapting news content in any way?
    \item Could you walk me through a specific example? How did it work?
    \item Who was the content intended for, and in what context would they encounter it?
\end{itemize}

\subsection{AI and Personalization in Existing Professional Practice}

\paragraph{AI in use.}
\begin{itemize}
    \item How, if at all, do you use AI tools in your work, either formally or experimentally?
    \item Which parts of the workflow does AI support (e.g., news gathering, production, verification, distribution, or moderation)?
    \item Could you walk me through a recent example?
\end{itemize}

\paragraph{News personalization.}
\begin{itemize}
    \item Who are the audiences you work with?
    \item Does personalization come up in your work?
    \item Are there situations where different audiences receive different content? How does that work?
\end{itemize}

\subsection{AI-Assisted News Personalization}

\paragraph{Hypothetical personalization scenario.}

Where relevant, participants were invited to consider a hypothetical tool that could adapt news for different readers
or platforms, for example by changing its length, tone, or focus.

\begin{itemize}
    \item What do you think of a tool like this?
    \item How might it change your current work?
    \item How much control would you want to keep over what it produces?
    \item What would you want readers to know about how a story had been adapted for them?
\end{itemize}

\paragraph{Editorial control and responsibility.}
\begin{itemize}
    \item How would responsibilities be divided if AI assisted with news adaptation or distribution?
    \item Which decisions would you want people to retain?
    \item How would you assess or investigate a problematic output?
    \item How would changes to a story or its sources be traced?
\end{itemize}

\subsection{Reflection and Closing}

\begin{itemize}
    \item Is there an experience from your work that you think is particularly relevant here?
    \item Is there anything important we haven't talked about?
    \item Is there anyone else you think we should speak to?
\end{itemize}

\section{Illustrative Coding-to-Finding Links}
\label{app:coding}

Table~\ref{tab:coding-overview} presents selected consolidated codes that informed the analytical findings. These examples illustrate connections between participants' accounts and our interpretations rather than an exhaustive or one-to-one mapping of codes to findings.

\begin{table*}[ht]
\centering
\caption{Illustrative links between consolidated codes and analytical findings. Code descriptions are condensed from the final coding scheme.}
\label{tab:coding-overview}
\small
\begin{tabularx}{\textwidth}{@{}p{0.21\textwidth}X p{0.27\textwidth}@{}}
\toprule
\textbf{Analytical finding} &
\textbf{Illustrative consolidated codes} &
\textbf{Analytical interpretation} \\
\midrule

Editorial boundaries:
form, selection, meaning
&
\texttt{INC3-045}: AI adaptation is acceptable when meaning and authorial intent are preserved.

\texttt{CTX3-009}: Shortening stories can remove context and introduce interpretation.
&
Changes in presentation can affect editorial meaning, complicating boundaries of delegation.
\\
\addlinespace

Tool personalization
&
\texttt{INC3-067}: Staff use personal or public AI tools outside approved systems.

\texttt{INC3-010}: Journalists edit AI-assisted drafts and remove recognizable AI style.
&
Individual tool adaptations become part of professional judgment and practice.
\\
\addlinespace

Different Framings of AI-Related Risks
&
\texttt{INC3-003}: AI produces unreliable facts, citations, and responses.

\texttt{INC3-058}: Personalized news raises concerns about filter bubbles and shared information.
&
Tool limitations and audience-facing personalization elicit different problem framings.
\\
\addlinespace

AI-mediated source horizons
&
\texttt{INC3-006}: AI misses local, linguistic, and platform-specific information.

\texttt{INC3-009}: AI supports brainstorming of new angles, questions, and experts.
&
AI-assisted research shapes which sources and perspectives become available for consideration.
\\
\addlinespace

Within-story informational consistency
&
\texttt{INC3-001}: Journalists trace AI-supplied facts and adapted versions to original sources.

\texttt{CTX3-009}: Shortening stories can remove context and introduce interpretation.
&
Different versions of a report may preserve or alter information necessary for interpretation.
\\
\addlinespace

Distributed control and accountability
&
\texttt{INC3-038}: Individuals are held responsible for AI-assisted outputs.

\texttt{INC3-039}: Managers introducing AI systems are accountable for their deployment.
&
Responsibility is negotiated across professional and organizational levels.
\\

\bottomrule
\end{tabularx}
\end{table*}

\section{Participant-Level Evidence Matrix}
\label{app:evidence-matrix}

Table~\ref{tab:matrix} summarizes how participants discussed tool and audience personalization, distinguished direct experience from other accounts, and framed associated risks. It documents the predominant pattern and qualifying cases discussed in Section~\ref{sec:experienced_anticipated}.

\begin{table}[ht]
\centering
\caption{Evidence matrix based on consolidated codes. \textbf{Tool} / \textbf{Aud.}: discussed tool or audience personalization ($\bullet$ from own practice; $\circ$ through observation, product work, or hypotheticals; -- not discussed). \textbf{Tool-limit framing}: Acc = accuracy or verification; Cap = capability or craft; Oth = other; $^{*}$ limitation experienced in own work. \textbf{Public resp.}: audience personalization linked to journalism's public responsibility ($\bullet$ explicit; $\circ$ indirect; -- not linked).}
\label{tab:matrix}
\scriptsize
\begin{tabular}{@{}lcclcp{0.5\linewidth}@{}}
\toprule
ID & Tool & Aud. & Tool-limit framing & Public resp. & Complicating or qualifying evidence \\
\midrule
P01 & $\circ$ & $\bullet$ & Acc & $\bullet$ & Expects falling hallucination rates to justify trust \\
P02 & $\bullet$ & $\bullet$ & Acc & $\bullet$ & -- \\
P03 & $\circ$ & $\bullet$ & Oth (low-value content) & $\bullet$ & Sees personalized versions widening inclusive access \\
P04 & $\bullet$ & $\bullet$ & Acc & $\bullet$ & Assigns recommendation to the publisher \\
P05 & $\circ$ & $\bullet$ & Cap$^{*}$ & $\bullet$ & ``Thin line'' between research support and framing information \\
P06 & $\circ$ & $\circ$ & Oth (voice, emotion) & -- & Expects audience tolerance of AI to grow \\
P07 & $\bullet$ & $\bullet$ & Cap, Acc$^{*}$ & $\bullet$ & Frames AI as broadening expert diversity; notes AI merging distinct statements \\
P08 & $\bullet$ & $\bullet$ & Acc$^{*}$ & $\bullet$ & -- \\
P09 & $\bullet$ & $\bullet$ & Acc$^{*}$ & $\bullet$ & Calls prompt-dependent answers ``bias'', treated as unreliability \\
P10 & $\circ$ & $\bullet$ & Acc & $\bullet$ & -- \\
P11 & $\bullet$ & $\bullet$ & Cap, Acc$^{*}$ & $\circ$ & Works around gaps with own language skills; links formats to public understanding \\
P12 & $\bullet$ & $\circ$ & Acc$^{*}$ & -- & Sees AI disrupting habitual thinking \\
P13 & $\bullet$ & $\bullet$ & Acc, Cap$^{*}$ & -- & Locates audience reach with platforms \\
P14 & $\bullet$ & $\circ$ & Acc, Cap$^{*}$ & -- & Commercial multi-version content for AI search \\
P15 & -- & $\circ$ & Acc$^{*}$ & $\circ$ & Limitations concern legal drafting; disclaims expertise on personalization \\
P16 & $\circ$ & $\circ$ & Oth (trust risk) & $\circ$ & Questions whether a single public sphere exists \\
\bottomrule
\end{tabular}
\end{table}

\section{Validation Studies}
\label{app:validation}
\subsection{AI-mediated source horizons}
\label{app:horizon}

\begin{figure*}[ht]
  \centering
  \includegraphics[width=\textwidth]{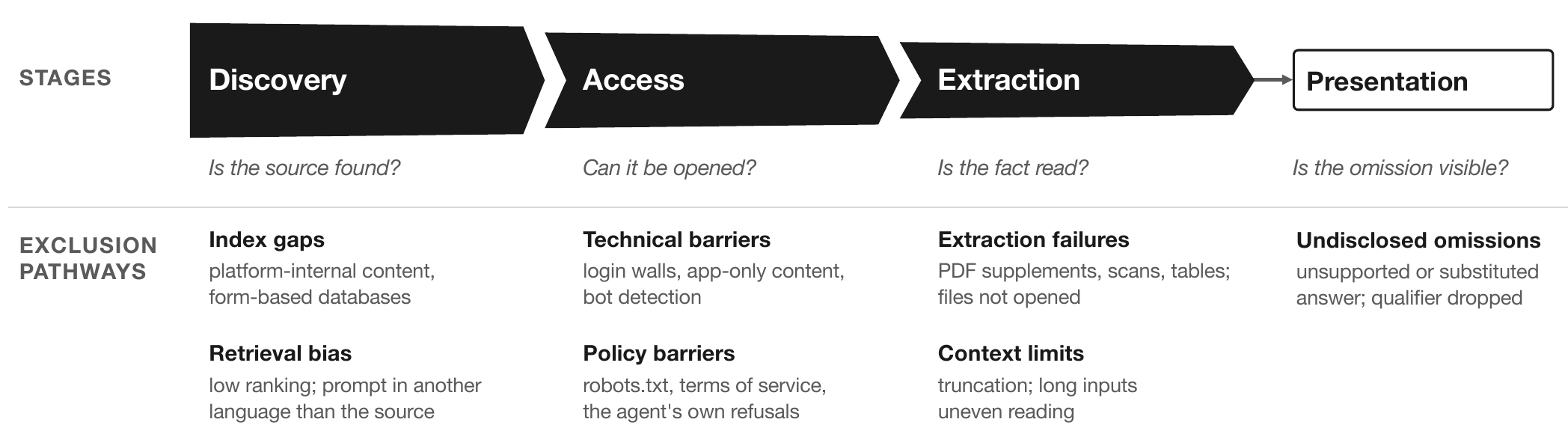}
  \caption{Information exclusion pathways in AI-assisted journalistic research. Three stages distinguish source discovery, accessibility, information extraction, and presentation. The pathways are informed by prior research and illustrated by four cases; they do not establish prevalence.}
  \Description{A three-stage diagram with progressively narrowing arrows labeled Discovery, Access, Extraction, and Presentation. Discovery includes index gaps and retrieval bias; Access includes technical and policy barriers; Extraction includes extraction failures and context limits; Presentation highlights undisclosed omissions. Examples include platform-hosted content, login restrictions, supplementary files, and missing qualifications.}
  \label{fig:source-horizon}
\end{figure*}

The AI-mediated source horizon concerns which sources and perspectives become available for journalistic consideration through AI-assisted research.
We distinguish three analytical stages through which information may be excluded from AI-assisted journalistic research: discovery, access, extraction, and presentation (Figure~\ref{fig:source-horizon}). 

\paragraph{Illustrative case design}

We examined four research tasks using three web-enabled AI systems: \texttt{GPT-6}, \texttt{Claude Sonnet 5.5}, and \texttt{Gemini 3}, accessed through their consumer interfaces on 31 October 2026. Two tasks compared an open question with the same question accompanied by a source URL; two others examined access to specified platform-hosted content. The illustrative cases show how these boundaries can arise in practice, without establishing their prevalence across systems.

We assessed whether the responses provided the requested information, acknowledged limitations, or offered unsupported alternatives. The analysis concerns observable responses, since internal retrieval and ranking decisions were not available. Complete prompts, source URLs, responses, and evaluation records are provided in the \href{https://anonymous.4open.science/r/genai-news-diversity/}{repository}.

\paragraph{Stage 1: Source discovery and prioritization}

The first stage concerns which sources become visible in AI-assisted research. Studies of generative search document concentrated citations among prominent news outlets~\citep{yang2025news}, unequal visibility across source categories~\citep{li2024arbiters,chen2025geo}, and language-dependent citation patterns~\citep{nguyen2026multilingual,carragher2026}. Controlled comparisons also show that models assign different weights to institutional and individual sources~\citep{schuster2026whose}.

These patterns arise from multiple selection mechanisms. Training data expose models to sources at different frequencies; Wikipedia, for example, has featured prominently in language-model training and influential retrieval benchmarks~\citep{brown2020gpt3,karpukhin2020dpr,petroni2021kilt}. Retrieval coverage, ranking, and answer generation further determine which sources appear. The resulting response presents selected information without necessarily revealing the alternatives.

\begin{quote}
\small
\noindent\textbf{Case A: An investigation absent from an open-query response.}

\noindent\textbf{Source:} A German-language CORRECTIV investigation concerning industrial groundwater extraction. (\href{https://correctiv.org/aktuelles/kampf-um-wasser/2022/11/22/klimawandel-wasser-knapp-industrie-hat-jahrzehntelange-entnahmerechte}{original article})

\noindent\textbf{Question:} Until what year is the B{\"o}klunder plant's groundwater permit valid, and what annual extraction volume does it allow?

\noindent\textbf{Open query:} \texttt{GPT-6} and \texttt{Claude Sonnet 5.5} reported that they could not locate the figures. \texttt{Gemini 3} supplied incorrect figures: 2040 and 450,000 cubic meters.

\noindent\textbf{URL supplied:} All three systems correctly reported 2042 and 1.3 million cubic meters per year.
\end{quote}

The source was publicly accessible, yet its information did not appear in responses to the open question. Providing the URL changed the outcome for every system. The distinction matters for journalism because the information available for editorial consideration depends partly on whether a practitioner relies on AI-led discovery or supplies independently identified sources.

\paragraph{Stage 2: Source accessibility}

A source can be identified yet remain inaccessible to an AI system. Login requirements, app-based publishing, anti-bot controls, and limited indexing constrain retrieval. Publishers also govern automated access through \texttt{robots.txt} and related policies~\citep{longpre2024consent,changhe2025}. Language and platform fragmentation further shape what systems can discover and retrieve~\citep{zhou2025browsecomp}.

\begin{quote}
\small
\noindent\textbf{Cases B and C: Information hosted on Facebook and WeChat.}

\noindent\textbf{Sources:}
A Brisbane City Council Facebook post about kerbside collection
(\href{https://www.facebook.com/BrisbaneCityCouncil/photos/a.446964827708/10157185562882709/?type=3}{original post})
and a WeChat article concerning a cooperation agreement
(\href{https://mp.weixin.qq.com/s/muM6WaMibGYqTO523al_5w}{original article}).

\noindent\textbf{Questions:} When was kerbside collection postponed, and which suburbs received the final scheduled collections? Where and when was the cooperation agreement signed, and by which three parties?

\noindent\textbf{URL supplied:} Five of six responses explicitly reported that the requested content could not be accessed. The remaining response, from \texttt{Gemini 3}, provided an uncited account of a collection suspension without establishing access to the Facebook post.

\noindent\textbf{Alternative information:} \texttt{Claude Sonnet 5.5} offered potentially related material for both tasks, explicitly stating that its connection to the requested source was unconfirmed.
\end{quote}

These cases reveal two forms of an access boundary: an acknowledged absence and a substitute answer. The latter can obscure the difference between information obtained from the requested source and information found elsewhere.

Access restrictions may serve legitimate purposes, including publishers' control over automated use of their content. Respecting these restrictions remains essential. Their diversity consequence is that information published through certain platforms or access arrangements can remain outside AI-assisted research, even when relevant to a reporting task.

\paragraph{Stage 3: Information extraction and presentation}

Even when relevant material is located, its substantive information may not reach the generated response. Tables, supplementary files, PDFs, and long documents introduce additional extraction and synthesis challenges. Research shows that models use information unevenly across long contexts~\citep{liu2024lost}, while generative search can produce statements whose citations provide incomplete evidential support~\citep{liu2023verifiability}.

\begin{quote}
\small
\noindent\textbf{Case D: A supplementary result omitted from the response.}

\noindent\textbf{Source:} A \href{https://weekly.chinacdc.cn/en/article/pdf/preview/dc8ccb25-9127-45d0-943d-424b7afa379c.pdf}{China CDC Weekly study} examining the June 2023 heat waves in northern China~\citep{chen2026heat}.

\noindent\textbf{Question:} How many excess deaths did the temperate-monsoon-zone sensitivity analysis estimate, including its 95\% confidence interval?

\noindent\textbf{Target information:} Supplementary Table~S2 reports 1,308 excess deaths (95\% CI: 754--1,945), compared with 1,036 in the main analysis.

\noindent\textbf{Open query:} All three systems referred to the main findings without supplying the supplementary estimate. \texttt{Gemini 3} identified Table~S2 but omitted its figures.

\noindent\textbf{URL supplied:} \texttt{GPT-6} returned the correct estimate. \texttt{Claude Sonnet 5.5} reported that it could not open the PDF; \texttt{Gemini 3} supplied a partial answer without the target estimate.
\end{quote}

The case distinguishes locating relevant evidence from conveying its contents. The systems identified the study, and one identified the relevant supplementary table, while none conveyed the requested estimate in their open-query responses.

For journalistic research, this distinction is consequential. Supplementary analyses, regional breakdowns, and alternative estimates can qualify headline findings. A response may point to the correct study while leaving out evidence essential to understanding what that study establishes.

\paragraph{From search mediation to AI-mediated sense-making}
\label{app:search-contrast}

Search engines have long shaped journalists' information environments. Indexing determines what can be found, ranking determines what receives attention~\citep{introna2000shaping}, and curated results concentrate exposure on selected outlets~\citep{trielli2019search}. Generative AI changes the locus of this mediation. Rather than presenting sources for journalists to examine, it increasingly selects, reads, extracts, and synthesizes information on their behalf~\citep{shah2022situating}. The journalist encounters an answer rather than the sources and decisions that produced it.

This shift changes how omissions become visible. In Case A, the CORRECTIV investigation was accessible through its URL but absent from open-query answers. Unlike a lower-ranked search result, an unselected source leaves no alternative to inspect in the synthesized response. In Cases A and B, incorrect or unsupported answers sometimes replaced missing information, making exclusion appear as completeness. In Case D, systems located the relevant study but omitted its supplementary estimate, illustrating how extraction determines which evidence reaches the journalist.

The interviews reveal how this mediation enters editorial judgment. P11 described how she ``used to search for all this information manually---typing queries one by one into the search bar on Google,'' an approach that ``resulted in information that was very scattered''; she now delegates this to AI tools that help her ``form a preliminary understanding of the story.'' AI thus mediates not only source discovery but the formation of initial interpretations. Practitioner configurations further shape this process through model choice, instructions, memory, and supplied materials (Section~\ref{sec:tool_personalization}).

The consequential shift is from ranking sources to constructing the account from which journalists begin their work. When discovery, reading, and synthesis occur inside the system, reviewing the answer reveals what was included, not what was excluded. The boundaries of journalistic consideration are shaped before editorial judgment begins.

\paragraph{Implications for news diversity}

AI-mediated sourcing shapes the information available for editorial judgment at multiple stages. Source discovery determines which materials are surfaced; accessibility determines which can be consulted; extraction and presentation determine which details reach the journalist. These stages are analytically distinct, although actual workflows may involve repeated searches, supplied documents, and independent sourcing.

Their implications extend beyond the factual accuracy of individual answers. Accuracy concerns whether the information presented is reliable; the AI-mediated source horizon also concerns what never becomes available for consideration. Sources outside that horizon can include less prominent reporting, information hosted on restricted platforms, and evidence embedded in supplementary materials.

A diversity perspective therefore asks which sources and perspectives become visible through AI-assisted research, what remains unseen, and how journalists account for those absences through their wider sourcing practices. The objective is to make informational boundaries recognizable and open to editorial judgment, while respecting legitimate restrictions on access.

\subsection{Within-story Informational Consistency}
\label{app:consistency}

We examined whether summarizing and adapting the same news article for different audiences preserves its main idea and the information essential to understanding it.

\paragraph{Design}
We selected 12 published news articles from Germany, Australia, and China, with two national and two local articles per country. Four generative models (\texttt{Qwen3-32B}, \texttt{gpt-6-luna}, \texttt{deepseek-flash}, and \texttt{gemini-3.8-flash}) adapted each article in its original language for five audiences (Table~\ref{tab:e2-audiences}). Generation was conducted through model APIs and locally hosted inference using vLLM.

Two prompt conditions compared a baseline adaptation instruction (P0) with an explicit instruction to preserve the source's meaning (P1). Each model generated one version per article, audience, and condition, yielding 480 adapted news versions in total. Comparing audience versions of the same article under the same prompt condition produced 960 matched audience pairs overall.

\begin{table}[ht]
\centering
\small
\caption{Audience specifications used in the consistency study, illustrated with the English prompts. German and Chinese specifications, including localized geographic and language-proficiency references, are provided in the \href{https://anonymous.4open.science/r/genai-news-diversity/}{repository}.}
\label{tab:e2-audiences}
\begin{tabular}{@{}lp{0.81\linewidth}@{}}
\toprule
ID & Target audience \\
\midrule
A1 & Young people aged 16--24 who mainly get their news on social media. \\
A2 & Listeners aged 65 and over who follow news on the radio; the text will be read aloud. \\
A3 & Readers whose first language is not English and who read English at CEFR B1 level. \\
A4 & Readers living in a small town in the Darling Downs region of Queensland. \\
A5 & Specialist readers of the relevant news desk who already know the subject well. \\
\bottomrule
\end{tabular}
\end{table}

All prompts specified a target length of 120--180 words for German and English, or 250--350 Chinese characters, informed by published editorial guidance on short-form news reporting~\citep{abaNewsGuidelines,chen2024efsa}.
The English prompt templates were:

\begin{quote}
\small
\noindent\textbf{P0: Baseline adaptation}

Adapt the following news article for \{audience\}. The adapted text should be \{length\_spec\}. Write in English. Return only the adapted text, without a headline or comments.

Article: \{article\_text\}
\end{quote}

\begin{quote}
\small
\noindent\textbf{P1: Explicit meaning preservation}

Adapt the following news article for \{audience\}. The adapted text should be \{length\_spec\}. Change only the form: length, wording, tone, examples and order. Do not change the facts, who said what, how certain statements are, who is responsible for what, or the balance between positions. Write in English. Return only the adapted text, without a headline or comments.

Article: \{article\_text\}
\end{quote}

For A5, the relevant desk was selected according to the article's subject (business, science and health, transport, environment, or culture and education). Complete multilingual prompt specifications are included in the \href{https://anonymous.4open.science/r/genai-news-diversity/}{repository}.

\paragraph{Evaluation}
We distinguished preservation of a story's main idea from retention of specific essential information.

\emph{Main-idea preservation.}
Using \texttt{gpt-oss-120b}, we assessed information insertion, deletion, and substitution following the factuality evaluation framework of \citet{devaraj2022factuality}. Each category was rated on a three-level severity scale: 0 (no or trivial change), 1 (nontrivial change that preserves the main idea), and 2 (change that alters the main idea).

\emph{Essential-information retention.}
Claude Opus 5.5 segmented the 12 articles into 342 information units. A researcher identified units essential to understanding each report, with selections cross-checked and resolved through discussion with a journalist, yielding 31 essential units (2--4 per article). Essentiality was assessed in context, considering core claims, substantively important numerical details, attributions, qualifications, and balancing perspectives.

We then estimated the retention of essential units using a multilingual \href{https://huggingface.co/MoritzLaurer/mDeBERTa-v3-base-xnli-multilingual-nli-2mil7}{mDeBERTa NLI model}. Each essential unit served as a hypothesis, tested against individual sentences and the full generated version as candidate premises. Inspired by the sentence-pair aggregation approach of SummaC-ZS~\citep{laban-etal-2022-summac}, we classified a unit as retained when the maximum entailment probability reached 0.4. Retention was the proportion of designated essential units classified as retained.

For each matched audience pair, we calculated the Jaccard distance between retained-unit sets:
\[
D(A,B)=1-\frac{|K_A\cap K_B|}{|K_A\cup K_B|},
\]
where $K_A$ and $K_B$ contain the essential units classified as retained. A distance of 0 indicates identical sets; 1 indicates no overlap. When both sets were empty, distance was assigned 0 by convention.

\emph{Validation and robustness checks.}
We manually reviewed 103 unit-level retention judgments to assess the NLI evaluator. At the initial threshold of 0.5, agreement with human judgments was 88.3\% (Cohen's $\kappa=0.765$). The researcher and journalist observed that disagreements primarily arose when the evaluator classified information as not retained despite human judgments to the contrary, suggesting an overly strict threshold. We therefore tested thresholds of 0.4 and 0.3. Both improved agreement to 91.3\%, with 0.4 yielding the highest Cohen's $\kappa$ (0.821; Table~\ref{tab:e2-robustness}). We used 0.4 for all subsequent analyses across the four models.

\begin{table}[ht]
\centering
\small
\caption{Agreement between NLI and human retention judgments across entailment thresholds ($n=103$).}
\label{tab:e2-robustness}
\begin{tabular}{@{}lcc@{}}
\toprule
NLI threshold & Agreement & Cohen's $\kappa$ \\
\midrule
0.5 & 88.3\% & 0.765 \\
0.4 (selected) & 91.3\% & 0.821 \\
0.3 & 91.3\% & 0.817 \\
\bottomrule
\end{tabular}
\end{table}

\paragraph{Findings}

\emph{AI adaptations preserved the main idea.}
Across four generators and 480 adaptations, none received a severity score of 2 for deletion, indicating that omissions did not alter the main idea of the original reporting. Human validation of 40 adaptations showed complete agreement with the automated judgments. These findings suggest that both commercial and open-weight models were generally effective at preserving the main idea in our adaptation tasks, even when some information was omitted.

\emph{Preserving the main idea did not ensure retention of essential details and qualifications.}
Mean essential-information retention ranged from 70.8\% to 80.3\% across models (Table~\ref{tab:e2-retention}). Explicit meaning-preservation instructions produced limited improvements of 2.8--8.9 percentage points. Even under P1, models retained only 72.2--82.8\% of essential information. Thus, explicitly instructing models to preserve meaning improved retention but did not prevent substantial omissions.

\begin{table}[ht]
\centering
\small
\caption{Mean essential-information retention across four generators at NLI threshold 0.4. Each model generated 60 versions per prompt condition. Differences are in percentage points (pp).}
\label{tab:e2-retention}
\begin{tabular}{@{}lrrrr@{}}
\toprule
Generator & P0 & P1 & Overall & P1--P0 \\
\midrule
GPT-6 Luna & 73.2\% & 82.1\% & 77.6\% & +8.9 pp \\
DeepSeek Flash & 77.8\% & 82.8\% & 80.3\% & +5.0 pp \\
Gemini 3.8 Flash & 73.6\% & 79.9\% & 76.7\% & +6.3 pp \\
Qwen3-32B & 69.4\% & 72.2\% & 70.8\% & +2.8 pp \\
\bottomrule
\end{tabular}
\end{table}

\emph{Audience versions differed in which essential information they retained.}
Across 960 matched audience pairs, mean Jaccard distance ranged from 0.212 for DeepSeek to 0.292 for Qwen, with 0.240 for GPT-6 Luna and 0.241 for Gemini (Table~\ref{tab:e2-pairs}). These distances capture differences in retained information rather than factual contradictions between versions.

The most divergent audience pairs varied across generators. For Qwen, young--rural versions showed the greatest divergence (0.410). For GPT-6 Luna, the largest difference was between older listeners and B1 readers (0.326); for DeepSeek, between B1 readers and rural readers (0.243); and for Gemini, between B1 readers and specialists (0.323). Thus, different audience specifications influenced information retention, but no single audience pairing was consistently the most divergent.
For example, in a German report on converting vacant buildings into affordable housing, DeepSeek's adaptation for young readers retained the policy's aims but omitted that safety requirements would remain in force, which was preserved in the version for B1-level readers. Both versions conveyed the same policy initiative, yet differed in whether readers were informed of a consequential safeguard.

\begin{table*}[t]
\centering
\small
\caption{Mean Jaccard distance between audience versions by generator at NLI threshold 0.4. Each audience comparison contains 24 matched pairs per generator. Bold indicates the highest value for each generator.}
\label{tab:e2-pairs}
\begin{tabular*}{0.7\textwidth}{@{\extracolsep{\fill}}lrrrr@{}}
\toprule
Audience pair & GPT & DeepSeek & Gemini & Qwen \\
\midrule
A1--A4 (young--rural)      & 0.250 & 0.205 & 0.274 & \textbf{0.410} \\
A2--A4 (older--rural)      & 0.201 & 0.233 & 0.274 & 0.368 \\
A3--A5 (B1--specialist)    & 0.288 & 0.222 & \textbf{0.323} & 0.250 \\
A1--A3 (young--B1)         & 0.236 & 0.184 & 0.236 & 0.368 \\
A2--A3 (older--B1)         & \textbf{0.326} & 0.163 & 0.153 & 0.264 \\
A4--A5 (rural--specialist) & 0.194 & 0.201 & 0.236 & 0.271 \\
A2--A5 (older--specialist) & 0.177 & 0.233 & 0.253 & 0.243 \\
A1--A2 (young--older)      & 0.181 & 0.222 & 0.125 & 0.257 \\
A1--A5 (young--specialist) & 0.236 & 0.212 & 0.260 & 0.264 \\
A3--A4 (B1--rural)         & 0.312 & \textbf{0.243} & 0.274 & 0.229 \\
\midrule
All pairs                  & 0.240 & 0.212 & 0.241 & 0.292 \\
\bottomrule
\end{tabular*}
\end{table*}

\emph{Interpretation.}
Retention measures the proportion of essential information preserved from the original article, while Jaccard distance measures how much audience-specific versions differ in which essential units they retain. A distance of 0 indicates identical retained information, while higher values indicate greater divergence, not necessarily factual contradictions.

Together, the results show that preserving a report's main idea does not guarantee retention of its essential information. Explicit meaning-preservation instructions improved retention but did not eliminate omissions or differences between audience versions.

\clearpage
\section*{Endmatter}

\subsection*{Generative AI Usage Statement}
\phantomsection
\label{sec:genai-usage}
All authors confirm the below uses of generative AI in this work.
We used Claude (Anthropic) to support the research process, not as an independent analyst or a substitute for researcher judgment.
Specifically, we used Claude in four ways.
First, as a brainstorming and ideation assistant during the development of the analytical framing.
Second, to help identify and retrieve related literature for comparison with our findings, however every suggested paper and claim was manually verified by the authors for relevance.
Third, to assist with consolidating and reducing the initial qualitative codes into the 141-code scheme described in Section~\ref{sec:analysis}; the authors reviewed the consolidated codes, examined 72 decisions flagged for attention, revised eight, and retained responsibility for all coding decisions. Interview transcripts processed with the tool had been pseudonymized beforehand.
Fourth, to support drafting and editing of the manuscript, including to cross-check quoted passages against interview transcripts and the coding workbook for wording.
The authors reviewed and revised all AI-assisted text and retain full responsibility for the manuscript contents.

\subsection*{Ethical Considerations and Adverse Impacts Statement}
This study received joint institutional ethics approvals from two universities where the author team are from, and all participants provided informed consent for voluntary, uncompensated interviews. Transcripts were pseudonymized for analysis, and identifying information was removed to protect participants, although combinations of professional roles, institutional contexts, and quotations may still pose risks of indirect identification.
Beyond participant confidentiality, our findings carry potential risks of misuse: recommendations for greater organizational oversight of AI-assisted journalism could be used to justify monitoring individual workflows or shifting responsibility for systemic failures onto front-line staff. Likewise, evaluating informational consistency across personalized news versions could encourage excessive standardization, undermining adaptations that improve accessibility for underserved audiences. We therefore distinguish accountability from workplace surveillance and the preservation of core information from uniformity of presentation. Finally, our purposive sample does not support generalizations across national media systems or organizations, and participants' anticipated audience harms should not be interpreted as demonstrated effects. The proposed evaluation and governance approaches are therefore presented as directions for further investigation, whose implementation must account for institutional resources, professional autonomy, and audience interests.

\end{document}